\documentclass[12pt]{article}

\usepackage[bitstream-charter]{mathdesign}
\usepackage[mathcal]{eucal}

\usepackage{float}
\usepackage{algorithm}
\usepackage{amsmath}
\usepackage{graphicx}
\usepackage{graphicx}
\usepackage{array}
\usepackage{tabularx}
\usepackage{color,epsfig}
\usepackage{cite}
\usepackage[caption = false,font = footnotesize]{subfig}
\usepackage{url}
\usepackage[colorlinks=true,citecolor=blue]{hyperref}
\usepackage{a4wide}
\usepackage[framemethod=tikz]{mdframed}

\definecolor{mycolor}{rgb}{0.122, 0.435, 0.698}

\newmdenv[innerlinewidth=0.5pt, roundcorner=4pt,linecolor=mycolor,innerleftmargin=6pt,
innerrightmargin=6pt,innertopmargin=6pt,innerbottommargin=6pt]{mybox}

\DeclareMathAlphabet{\baz}{OML}{cmm}{b}{i}
\def\x{{\mathbf x}}

\def\bW{\mathbf{W}}

\def\b0{\mbox{\boldmath $0$}}

\def\bx{\mathbf{x}}
\def\by{\mathbf y}
\def\bz{\mathbf{z}}

\def\x{{\mathbf x}}

\def\x{{\mathbf x}}

\newtheorem{theorem}{Theorem}
\newtheorem{proposition}[theorem]{Proposition}

\floatstyle{ruled}
\newfloat{algorithm}{tbp}{loa}
\providecommand{\algorithmname}{Algorithm}
\floatname{algorithm}{\protect\algorithmname}

\begin{document}

\title{Distributed Stochastic Nonconvex Optimization and Learning based on Successive Convex Approximation}

\author{Paolo~Di Lorenzo and Simone Scardapane\thanks{Authors are with the Department of Information Engineering, Electronics, and Telecommunications, Sapienza University of Rome, Via Eudossiana 18, 00184, Rome, Italy.
E-mail: \{paolo.dilorenzo,simone.scardapane\}@uniroma1.it.\ This work was funded by Sapienza project n. RP11816431B9E1BC.}}

\maketitle

\begin{mybox}
\textbf{Note}: This paper was presented at the 2019 Asilomar Conference on Signals, Systems, and Computers. Due to a miscommunication, the paper was mistakenly omitted from the original submission to IEEE Xplore. It is now included as part of the conference record.
\end{mybox}

\begin{abstract}
We study \emph{distributed stochastic nonconvex} optimization in multi-agent networks. We introduce a novel algorithmic framework for the distributed minimization of the sum of the expected value of a smooth (possibly \emph{nonconvex}) function--the agents' sum-utility--plus a convex (possibly nonsmooth) regularizer. The proposed method hinges on successive convex approximation (SCA) techniques, leveraging \emph{dynamic consensus} as a mechanism to track the average gradient among the agents, and recursive averaging to recover the expected gradient of the sum-utility function. Almost sure convergence to (stationary) solutions of the nonconvex problem is established. Finally, the method is applied to distributed stochastic training of neural networks. Numerical results confirm the theoretical claims, and illustrate the advantages of the proposed method with respect to other methods available in the literature. 
\end{abstract}

%\vspace{.1cm}
%\begin{IEEEkeywords}
%Distributed optimization, nonconvex, %stochastic optimization,  successive %convex approximation, consensus.
%\end{IEEEkeywords}

\section{Introduction}

Recent years have witnessed a surge of interest in distributed optimization methods for multi-agent systems. In the stochastic setting, many such problems can be formulated as the cooperative minimization of the expected agents' sum-utility $F$ plus a regularizer $G$:
\begin{align}\label{Problem}
& \underset{\bx}{\text{minimize}} \;\;  \mathbb{E}[ F(\bx,\boldsymbol{\xi})]+G(\bx)\\
& \text{subject to} \;\; \bx\in \mathcal{K},\nonumber
\end{align}
where 
$F(\bx,\boldsymbol{\xi})\triangleq \sum_{i=1}^{I}f_i(\bx,\boldsymbol{\xi})$ 
is the sum-utility function, with each $f_i(\bx,\boldsymbol{\xi})$ being the smooth (possibly \emph{nonconvex}, nonseparable) cost function of agent $i\in \{1,\ldots, I\}$ that depends on the variable $\bx \in \mathbb{R}^p$ and a random vector $\boldsymbol{\xi}$, whose probability distribution is defined on $\mathcal{D}\subseteq \mathbb{R}^p$; $G$ is a  convex (possibly nonsmooth,  nonseparable) function; and $\mathcal{K}\subseteq \mathbb{R}^p$ is closed and convex. Usually the nonsmooth term is used to promote some extra structure in the solution, typically sparsity.

Network-structured optimization problems in the form (\ref{Problem}) are found widely in several engineering areas, including sensor networks information processing, communication networks, multi-agent control and coordination, and distributed machine learning, just to name a few. Common to these problems is the necessity of performing a  decentralized computation/optimization, due to the large size of the network and volume of data,  energy constraints, and/or privacy issues. Motivated by these observations, this paper aims to develop a provable solution method for the general class of \emph{nonconvex stochastic}  problems (\ref{Problem}), in the following distributed setting: i) the network of agents is modeled as a directed (strongly connected) graph;  ii) agents  know their local functions $f_i$ only, the common regularizer $G$, and the feasible set $\mathcal K$; and iii) only communications between single-hop neighbors are possible. 

\noindent \textbf{Related works.} Distributed solution methods for \emph{convex} and \emph{deterministic} instances of Problem (\ref{Problem}) have been widely studied in the literature; they are usually either primal (sub)gradient-based methods \cite{Nedic-Ozdaglar,Nedic-Ozdaglar-Parillo,jakovetic2014fast}, or primal-dual schemes, e.g.,  \cite{Boyd-Parikh-Chu-Peleato-Eckstein}. Similarly, distributed strategies for \emph{convex} and \emph{stochastic} instances of (\ref{Problem}) are either diffusion adaptation schemes \cite{Cattivelli-Sayed,Chen-Sayed,DiLo-Sayed,vlaski2016diffusion}, or ADMM algorithms \cite{ouyang2013stochastic,zhong2014fast}. The literature on distributed \emph{nonconvex} optimization is much more recent. The \emph{nonconvex} and \emph{deterministic} setting includes: i) primal gradient-based methods \cite{zeng2018nonconvex}; ii) Frank-Wolfe algorithms \cite{wai2017decentralized}; iii) SCA methods \cite{di2016next}; proximal primal-dual algorithms \cite{hong2017prox}; and distributed annealing schemes \cite{swenson2019annealing}.
Finally, \emph{nonconvex} and \emph{stochastic} instances of (\ref{Problem}) have been considered only very recently in \cite{vlaski2019distributed,vlaski2019distributed2,vlaski2019second,george2019distributed,lu2019gnsd}. In particular, the works in \cite{vlaski2019distributed,vlaski2019distributed2} illustrate how distributed stochastic gradient algorithms achieves agreement at linear rate while escaping saddle points. The work in \cite{george2019distributed} provide sufficient conditions to guarantee asymptotic mean-square convergence of distributed stochastic gradient methods, considering twice differentiable objective functions. Finally, the work in \cite{lu2019gnsd} propose a first-order distributed algorithm based on gradient-tracking, which finds stationary points with guaranteed convergence rate. 

\noindent \textbf{Contributions.} All previous art on distributed stochastic nonconvex optimization is based on first-order methods that exploit only gradient information of the objective functions $f_i$, and does not consider constraints. This paper introduces the first distributed (best-response-based) algorithmic framework for the \emph{distributed}, \emph{stochastic}, \emph{nonconvex}, \emph{constrained} optimization in the general form (\ref{Problem}). The crux of the framework is a \emph{convexification-decomposition} technique that hinges on SCA methods \cite{Scutari-Facchinei-Song-Palomar-Pang,di2016next}, while leveraging \emph{dynamic} consensus as a gradient tracking mechanism, and recursive average to asymptotically recover the gradient of the expected loss function; we will term it as Stochastic in-Network succEssive conveX approximaTion algorithm (S-NEXT). Almost sure convergence to (stationary) solutions of the nonconvex problem (\ref{Problem}) is established. Numerical simulations on distributed stochastic training of neural network models confirm the theoretical results, and assess the performance of the proposed method over real datasets.

\section{In-Network Stochastic Nonconvex Optimization via SCA}

Consider a  network composed of $I$ autonomous agents  aiming to cooperatively and distributively solve   Problem (\ref{Problem}).\smallskip

\noindent \textbf{Assumption A.} We make the following blanket assumptions:
\begin{description}
\item[(A1)]  The set $\mathcal{K}$ is (nonempty) closed and convex;
\item[(A2)] Each $f_i$ is $C^1$ (possibly nonconvex) on $\mathcal K$;
\item[(A3)]  $\nabla f_i$ is Lipschitz continuous and bounded on $\mathcal{K}$;
\item[(A4)] $G$ is a convex function (possibly nondifferentiable) with bounded subgradient on $\mathcal{K}$;
\item[(A5)] $U$ is coercive; 
\item[(A6)] $\boldsymbol{\xi}$ is a bounded i.i.d. random vector defined on set $\mathcal{D}$.
\end{description}

Assumptions above are standard and  satisfied by many practical problems. Note that $f_i$'s need not be convex. In the following, we also make the blanket assumption that each agent $i$ knows only its own $f_i$ (but not $F$), the common $G$, and the feasible set $\mathcal K$. \\
\noindent \texttt{On network topology:}   The network of the agents is modeled as a directed graph $\mathcal{G}=(\mathcal{V,E})$, where $\mathcal{V}=\{1,\ldots,I\}$ is the vertex (i.e., agent) set, and $\mathcal{E}$ is the set of edges. The neighborhood of agent $i$ (including node $i$) is defined as
$\mathcal{N}_i=\{j|(j,i)\in\mathcal{E}\}\cup\{i\}$; it sets the communication pattern between single-hop neighbors: agents $j \neq i$ in $\mathcal{N}_i$ can communicate with node $i$.
We introduce the weights $w_{ij}$ matching the graph $\mathcal{G}$, i.e. $w_{ij}>0$ if $j\in \mathcal{N}_i$.
We also define the matrix $\bW\triangleq (w_{ij})_{i,j=1}^I$. We make the following weak assumptions on the network connectivity.

\vspace{0.5em}
\noindent{\bf  (A7)} \, The graph $\mathcal{G}$ is connected. Furthermore, the weight matrix $\bW$  satisfies
$\bW\,\mathbf{1}=\mathbf{1}$ and $\mathbf{1}^T \bW=\mathbf{1}^T$.
\vspace{0.5em}

Our goal is to develop an algorithm that converges to stationary solutions of Problem (\ref{Problem}) while being implementable  in the above distributed setting. To shed light on the core idea of our decomposition technique, we introduce first an informal and constructive description of the proposed scheme.

\subsection{Development of S-NEXT: A constructive approach}

Devising  distributed solution methods for Problem (\ref{Problem}) faces three main challenges, namely: the impossibility to evaluate the expectation accurately (e.g., because the statistics of the random
variables are unknown and/or the computational complexity is
prohibitive), the nonconvexity of $F$, and the lack of global information on $F$. To cope with these issues, we propose to combine SCA techniques  (Step 1 below), recursive averaging (Step 2), and dynamic consensus mechanisms (Steps 3 and 4), as described next. 

\noindent \texttt{Step 1 (local SCA optimization):} Each agent $i$ maintains a local estimate $\bx_i$ of the optimization vector $\bx$ that is iteratively updated. Solving directly Problem (\ref{Problem}) may be too costly (due to the expectation and the nonconvexity of $F$) and is not even doable in a distributed setting (because of the lack of knowledge of the whole $F$). One may then prefer to approximate Problem (\ref{Problem}), in some suitable sense, in order
to permit each agent to compute \emph{locally} and \emph{efficiently} the new iteration. Thus, to handle the nonconvexity of $F$ at every iteration $t$,  given the local  estimate $\bx_i^t$, each agent $i$ should solve the following \emph{strongly convex} optimization problem:
\begin{equation}\label{best_resp_x_hat}
\widehat{\bx}_i^t\triangleq{\rm arg\!}\min_{\!\!\!\!\!\!\!\!\!\!\!\!\!\boldsymbol{\bx_i}\in\mathcal{K}}\; \mathbb{E}\Big[ \widetilde{F}_i(\bx_i;\bx_i^t,\boldsymbol{\xi})\Big]+G(\bx_i),\vspace{-.1cm}
\end{equation}
where $\widetilde{F}_i(\bx_i;\bx_i^t,\boldsymbol{\xi})$ is a suitably chosen \emph{strongly convex} surrogate  of the nonconvex original ${F}(\bx,\boldsymbol{\xi})$, which may depend on the current iterate $\bx_i^t$. The main idea behind  \eqref{best_resp_x_hat} is to compute stationary solutions of Problem (\ref{Problem}) as fixed-points of the mappings $\widehat{\bx}_i(\bullet)$. The next proposition addresses the  question   about the connection  between such   fixed-points and stationary solution; its proof follows the same steps as \cite[Prop. 8(b)]{Scutari-Facchinei-Sagratella} and thus is omitted.

\begin{proposition}\label{Prop:fixed-point-stationary} Given Problem (\ref{Problem}) under A1-A6, suppose that  $\widetilde{F}_i$ satisfies the following conditions:
 \begin{description}
\item[ (F1)]
 $\widetilde{F}_{i} (\mathbf{\bullet}; \mathbf{y},\boldsymbol{\xi})$ is uniformly strongly convex on $\mathcal K$;\smallskip
\item[  (F2)]  $\nabla \widetilde{F}_{i} (\mathbf{x};\mathbf{x},,\boldsymbol{\xi}) = \nabla F(\mathbf{x},\boldsymbol{\xi})$ for all $\mathbf{x} \in \mathcal K$, $\boldsymbol{\xi}\in \mathcal D$;\smallskip
\item[  (F3)]  $\nabla \widetilde{F}_{i} (\mathbf{x};\mathbf{\bullet},\boldsymbol{\xi})$ is uniformly Lipschitz continuous
on $ \mathcal K$.
\end{description}
Then, the set of fixed-point of $\widehat{\bx}_i(\bullet)$ coincides with that of the stationary solutions of  (\ref{Problem}). Therefore, $\widehat{\bx}_i(\bullet)$ has a fixed-point. \
\end{proposition}

Conditions F1-F3 are quite natural:  $\widetilde{F}_{i} $ should be regarded
as a (simple) convex, approximation of $F$ at the point $\bx$ that preserves the first order properties of $F$.

\noindent \texttt{Step 2 (Recursive Averaging):} The issue with \eqref{best_resp_x_hat} is that usually the expectation cannot be computed in closed form. To deal with it, we follow the approach proposed in \cite{yang2016parallel}. Thus, given the realization of the random variable $\boldsymbol{\xi}$ at time $t$, i.e., $\boldsymbol{\xi}^t\in\mathcal{D}$, we propose to build the sample approximation of $\mathbb{E} [\widetilde{F}_i(\mathbf{x}_i;\mathbf{x}_i^t,\boldsymbol{\xi})]$ as:
\begin{equation}\label{sample_approx}
\overline{F}_i(\mathbf{x}_i;\mathbf{x}_i^t,\boldsymbol{\xi}^t) = \rho^t \,\widetilde{F}_i(\mathbf{x}_i;\mathbf{x}_i^t,\boldsymbol{\xi}^t) + (1-\rho^t) {\mathbf{d}_i^t}^T \left( \mathbf{x}_i-\mathbf{x}_i^t\right)
\end{equation}
where $\rho^t$ is a suitably chosen step-size sequence, and $\mathbf{d}_i^t$ is an online estimate of the gradient of ${\mathbb{E} [F(\mathbf{x}_i,\boldsymbol{\xi})]}$ that is recursively updated as:
\begin{equation}\label{di_recursion}
    \mathbf{d}_i^{t+1} = \left(1-\rho^t\right)\mathbf{d}_i^t + \rho^t \,\nabla F(\mathbf{x}_i^t,\boldsymbol{\xi}^t).
\end{equation}
Then, using (\ref{sample_approx}) in (\ref{best_resp_x_hat}), each node $i$ at time $t$ solves the following \emph{strongly convex} optimization problem:
\begin{align}\label{best_resp_x_hat2}
\widehat{\bx}_i^t\triangleq{\rm arg\!}\min_{\!\!\!\!\!\!\!\!\!\!\!\!\!\boldsymbol{\bx_i}\in\mathcal{K}}\; \Big\{\rho^t \widetilde{F}_i(\mathbf{x}_i;\mathbf{x}_i^t,\boldsymbol{\xi}^t) + (1-\rho^t) {\mathbf{d}_i^t}^T \left( \mathbf{x}_i-\mathbf{x}_i^t\right)+G(\bx_i)\Big\},
\end{align}
where $\mathbf{d}_i^t$ is updated as in (\ref{di_recursion}).

\noindent \texttt{Step 3 (Gradient Tracking):} The SCA step in \eqref{best_resp_x_hat2} handles the expectation and the nonconvexity in (\ref{Problem}). Nevertheless, \eqref{best_resp_x_hat2} still cannot be computed locally by node $i$ because of the lack of global information needed to build $\widetilde{F}_i(\mathbf{x}_i;\mathbf{x}_i^t,\boldsymbol{\xi}^t)$, and to update $\mathbf{d}_i^t$ in (\ref{di_recursion}).  To cope with the first issue (i.e., the choice of $\widetilde{F}_i$), since node $i$ has knowledge only of $f_i$, writing  $F(\bx_i,\boldsymbol{\xi})=f_i(\bx_i,\boldsymbol{\xi})+\sum_{j\neq i}f_j(\bx_i,\boldsymbol{\xi})$, leads naturally to a $\widetilde{F}_i$ wherein the (possibly) nonconvex  $f_i(\bx_i,\boldsymbol{\xi})$ is replaced by a convex surrogate $\widetilde{f}_i(\bx_i;\bx_i^t,\boldsymbol{\xi})$ and  $\sum_{j\neq i}f_j(\bx_i,\boldsymbol{\xi})$  is  linearized around the current iterate $\bx_i^t$. More formally, each agent $i$ solves the subproblem: given $\bx_i^t$,
\begin{align}\label{best_resp_x_hat3}
\widehat{\bx}_i^t\triangleq{\rm arg\!}\min_{\!\!\!\!\!\!\!\!\!\!\!\!\!\boldsymbol{\bx_i}\in\mathcal{K}}\; \Big\{\rho^t \Big(  \underset{\widetilde{F}_{i}(\bx_{i};\bx_{i}^t,\boldsymbol{\xi}^t)}{\underbrace{ \widetilde{f}_{i}(\bx_{i};\bx_{i}^t,\boldsymbol{\xi}^t)+\boldsymbol{\pi}_{i}(\bx_{i}^t,\boldsymbol{\xi}^t)^{T}(\bx_{i}-\bx_{i}^t)}} \Big)+ (1-\rho^t) {\mathbf{d}_i^t}^T \left( \mathbf{x}_i-\mathbf{x}_i^t\right)+G(\bx_i)\Big\},
\end{align}
where  
\begin{equation}\label{pi}
    \boldsymbol{\pi}_i(\bx_i^t,\boldsymbol{\xi}^t)\triangleq\sum_{j\neq i}\nabla_{\bx}f_j(\bx_i^t,\boldsymbol{\xi}^t).
\end{equation}
It is easy to check that $\widetilde{F}_i$ in \eqref{best_resp_x_hat3}  satisfies F1-F3 if also $\widetilde{f}_{i}$ satisfies them. An appropriate choice of $\widetilde{f}_i$ depends on the problem at hand and on computational requirement. The computation of $\widehat{\bx}_i^t$ in \eqref{best_resp_x_hat3} is still not fully distributed, because the evaluation of
$\boldsymbol{\pi}_i(\bx_i^t,\boldsymbol{\xi}^t)$ in (\ref{pi}) and the update of $\mathbf{d}_i^t$ in (\ref{di_recursion}) would require the knowledge of all $\nabla f_j(\bx_i^t)$, which is not available locally at node $i$. This lack of global knowledge can be solved exploiting \emph{dynamic} average consensus methods \cite{Zhu-Martinez}, which enable to track the network average gradient via local exchange of information between neighbors, as proposed in \cite{di2016next}. In particular, letting $\by_i^t$ be the local estimate at agent $i$ for $\overline{\nabla f}(\bx_i^t,\boldsymbol{\xi}^t)=(1/I)\sum_{j=1}^I\nabla_{\bx}f_j(\bx_i^t,\boldsymbol{\xi}^t)$, this can be done updating $\by_i^t$ according to:\vspace{-0.1cm}
\begin{equation}\label{y2}
\by_i^{t}\triangleq\sum_{j=1}^I w_{ij}\by_j^{t-1}  + \nabla f_i(\x_i^{t},\boldsymbol{\xi}^t)-\nabla f_i(\bx_i^{t-1},\boldsymbol{\xi}^{t-1})
\end{equation}
with $\by_i^0\triangleq\nabla f_i(\bx_i^0,,\boldsymbol{\xi}^0)$. Thus, given $\by_i^{t}$ in (\ref{y2}), the local estimates at node $i$ for $\boldsymbol{\pi}_i(\bx_i^t,\boldsymbol{\xi}^t)$ in (\ref{pi}) and for $\nabla F(\mathbf{x}_i^t,\boldsymbol{\xi}^t)$ in (\ref{di_recursion}) are given by:
\begin{align}\label{local_estimates}
    \widetilde{\boldsymbol{\pi}}_i^t = I\by_i^t-\nabla,  f_i(\x_i^{t},\boldsymbol{\xi}^t),  \qquad
    \widetilde{\nabla F}(\mathbf{x}_i^t,\boldsymbol{\xi}^t)= I \by_i^t,
\end{align}
respectively. Note that since the weights $w_{ij}$ are constrained by the network topology, the update of $\by_i^t$ in (\ref{y2}), and thus $\widetilde{\boldsymbol{\pi}}_i^t$ and $\widetilde{\nabla F}(\mathbf{x}_i^t,\boldsymbol{\xi}^t)$ in (\ref{local_estimates}), can be now performed locally with message exchanges with the agents in the neighborhood $\mathcal N_i$.

\noindent \texttt{Step 4 (Agreement):} To force the asymptotic agreement among the $\bx_i$'s,  a consensus-based step is employed on $\widehat{\bx}_i^t$'s. Each agent $i$ updates its  $\bx_i$ as:%by combining the   solutions received   from its neighbors:
 \begin{equation}\label{consensus_update}
\bx_i^{t+1}= \sum_{j=1}^I w_{ij}\, \widehat{\bx}_j(\bx_j^{t}),\vspace{-0.1cm}
\end{equation}
which can be implemented via local message exchanges in each node's neighborhood.

\subsection{The S-NEXT algorithm}

We are now in the position to formally introduce the S-NEXT algorithm, Algorithm 1; its convergence to stationary solutions of Problem (\ref{Problem}) is stated in Theorem \ref{simplified_convergence_th}, whose proof is omitted because of space limitations. S-NEXT algorithm builds on the iterates \eqref{best_resp_x_hat3} (wherein $\boldsymbol{\pi}_i(\bx_i^t,\boldsymbol{\xi}^t)$ is replaced by $\widetilde{\boldsymbol{\pi}}_i^t$ in (\ref{local_estimates})), (\ref{di_recursion}) (wherein $\nabla F(\mathbf{x}_i^t,\boldsymbol{\xi}^t)$ is replaced by $\widetilde{\nabla F}(\mathbf{x}_i^t,\boldsymbol{\xi}^t)$ in (\ref{local_estimates})),  \eqref{y2} and \eqref{consensus_update} introduced in the previous section. Also, in S1, in addition to solving the strongly convex optimization problem (\ref{best_resp_x_hat3}), we also introduced a step-size sequence $\alpha^t$ in the iterate: the new point $\bz_i^t$ is a convex combination
of  the current estimate ${\bx}_i^t$ and the solutions of (\ref{best_resp_x_hat3}). The convergence properties of S-NEXT are illustrated in the following Theorem.\smallskip

\begin{theorem}\label{simplified_convergence_th}
Given Problem (\ref{Problem}) under A1-A7, let $\{\mathbf{x}^t\}_n\triangleq \{(\mathbf{x}_i^t)_{i=1}^I\}_n$ be the sequence generated by Algorithm 1, and let $\{\overline{\mathbf{x}}^t\}_t\triangleq \{(1/I)\sum_{i=1}^I\mathbf{x}_i^t\}_t$ be its average. Choose the step-size sequences $\{\alpha^t\}_t$ and $\{\rho^t\}_t$ so that: 
\begin{itemize}
\item $\alpha^t\in (0,1]$ $\forall t\;$, $\; \sum_{t=0}^{\infty}\alpha^t=
\infty\;$, and $\;\sum_{t=0}^{\infty}(\alpha^t)^2<\infty$;\smallskip
\item $\rho^t\in (0,1]$ $\forall t\;$, $\;\sum_{t=0}^{\infty}\rho^t=
\infty\;$, and $\;\sum_{t=0}^{\infty}(\rho^t)^2<\infty$;\smallskip
\item $\displaystyle \lim _{t\rightarrow \infty} \alpha^t/\rho^t =0$. \smallskip
\end{itemize}
Then, we have:
\\(a) \emph{\texttt{[convergence]}:} the sequence $\{\overline{\mathbf{x}}^t\}_t$ is bounded and all its limit points are stationary solutions of (\ref{Problem}) almost surely; \\(b) \emph{\texttt{[consensus]}}:   all the sequences  $\{\bx_i^t\}_t$  asymptotically agree, i.e., $\|\mathbf{x}_{i}^t-\overline{{\mathbf{x}}}^t\|\underset{t\rightarrow\infty}{\longrightarrow}0
 $, for all $i=1,\ldots ,I$.
\end{theorem}

\begin{algorithm}[t]
\caption{: Stochastic In-Network Nonconvex Optimization}
\textbf{Data:} $\alpha^t,\rho^t>0$, $\mathbf{x}_{i}^0\in\mathcal{K}$, $\mathbf{y}_{i}^0=\nabla f_{i}(\mathbf{x}_{i}^0,\boldsymbol{\xi}^0)$, $\widetilde{\boldsymbol{{\pi}}}_{i}^{0}=(I-1)\mathbf{y}_{i}^0$, $\mathbf{d}_{i}^0=I\cdot \mathbf{y}_{i}^0$, for $i\in\mathcal{V}$; Set $t=0$;\smallskip

\textbf{(S1) SCA Optimization:} Each agent $i$ evaluates:
\begin{gather}
{\color{black}\widehat{\mathbf{x}}_{i}^t}=\underset{\color{black}\mathbf{x}_{i}\in\mathcal{K}}{\text{argmin}}
\Big\{
\rho^t\Big(\widetilde{f}_{i}\left(\mathbf{x}_{i};\mathbf{x}_{i}^t,\boldsymbol{\xi}^t\right)
+\widetilde{{\boldsymbol{{\pi}}}}_{i}^t\;^{T}(\mathbf{x}_{i}-\mathbf{x}_{i}^t)
\Big) + (1-\rho^t)  {\mathbf{d}_i^t}^T \left( \mathbf{x}_i-\mathbf{x}_i^t\right)+G(\mathbf{x}_{i})
\Big\} \nonumber\\
\hspace{-2cm}{\color{blue}{\color{black}\mathbf{z}_{i}^t=\mathbf{x}_{i}^t+\alpha^t\left(\widehat{\mathbf{x}}_{i}^t-\mathbf{x}_{i}^t\right)}} \nonumber
\end{gather}

\textbf{(S2) Agreement and Gradient Tracking:} Each agent $i$ collects data from its neighbors and updates the variables
$\mathbf{x}_{i}^t$, $\mathbf{y}_{i}^t$, and $\widetilde{{\boldsymbol{{\pi}}}}_{i}^t$ as:
\begin{align}
&\mathbf{x}_{i}^{t+1} = {\displaystyle {\sum_{j\in\mathcal{N}_{i}^t}}}\!\!w_{ij}^t\,\mathbf{z}_{j}^t\smallskip \nonumber\\
& \mathbf{y}_{i}^{t+1} = {\displaystyle {\sum_{j\in\mathcal{N}_{i}^t}}}\!w_{ij}^t\,\mathbf{y}_{j}^t+\nabla f_{i}(\mathbf{x}_{i}^{t+1},\boldsymbol{\xi}^{t+1})\hspace{-0.02cm}-\hspace{-0.02cm}\nabla f_{i}(\mathbf{x}_{i}^{t},\boldsymbol{\xi}^{t})\smallskip\nonumber\\
&\widetilde{\boldsymbol{{\pi}}}_{i}^{t+1} = I\cdot\mathbf{y}_{i}^{t+1}-\nabla f_{i}(\mathbf{x}_{i}^{t+1},\boldsymbol{\xi}^{t+1})\nonumber
\end{align}

\textbf{(S3) Gradient Averaging:} Each agent $i$ updates the local variable $\mathbf{d}_i^{t}$ as:
\begin{align}
\mathbf{d}_i^{t+1} =  \left(1-\rho^t\right)\mathbf{d}_i^{t} + \rho^{t} I \cdot\mathbf{y}_{i}^{t+1}\nonumber
\end{align}

\textbf{(S4)} If $(\mathbf{x}_{i}[n])_{i}$ satisfies a termination rule, STOP; otherwise, $t\leftarrow t+1$ and go to (S.1).
\end{algorithm}

\section{Application to Distributed Stochastic Training of Neural Networks}

As  a specific application of the S-NEXT framework, we consider the distributed training of neural network (NN) models, a problem of significant practical interest \cite{scardapane2017framework,wen2017terngrad}. Let us then assume a scenario where $I$ agents collect input-output pairs $(y_{i,m},\mathbf{x}_{i,m})$, for  $m\in {\cal S}_i$, $i=1,\ldots,N$. Also, let us denote by $g(\mathbf{w},\mathbf{x})$ a generic neural network with weight vector parameter $\mathbf{w}$, and taking $\mathbf{x}$ as input. Then, the distributed training problem can be mathematically cast as \cite{scardapane2017framework}:
\begin{equation}\label{NN_training}
    \underset{\mathbf{w}}{\text{min}} \;\;\; \displaystyle\sum_{i=1}^{I}
\underbrace{\frac{1}{|{\cal S}_i|}\sum_{m\in {\cal S}_i} l\left(y_{i,m},g(\mathbf{w},\mathbf{x}_{i,m})\right)}_{f_i(\mathbf{w})}+G(\mathbf{w}),
\end{equation}
where $l$ is a convex loss function (e.g., squared loss, cross-entropy, etc.), and $G$ is a convex regularizer (e.g., the $\ell_2$ norm). When the size of the dataset becomes very large, direct optimization of (\ref{NN_training}) might become prohibitive. A common approach to reduce complexity is to draw random mini-batch of data at every iteration, say, $\mathcal{B}_i^t\subseteq{\cal S}_i$, in order to approximate the global cost in (\ref{NN_training}). This approach leads to the following stochastic optimization problem:
\begin{equation}\label{NN_training_stochastic}
\underset{\mathbf{w}}{\text{min}} \;\;\; \displaystyle  \mathbb{E}\Bigg[\sum_{i=1}^{I}\underbrace{\frac{1}{|{\cal B}_i^t|}\sum_{m\in {\cal B}_i^t} l\left(y_{i,m},g(\mathbf{w},\mathbf{x}_{i,m})\right)}_{f_i(\mathbf{w},\boldsymbol{\xi})}\Bigg]+G(\mathbf{w}),
\end{equation}
which we aim to solve in a distributed fashion using the S-NEXT framework. Indeed, each function $f_i(\mathbf{w},\boldsymbol{\xi})$ in (\ref{NN_training_stochastic}) is nonconvex beecause of the presence of the NN function $g$, and is dependent on a random parameter $\boldsymbol{\xi}$ that models the random data sampling at each iteration. Our proposed approach is to use the S-NEXT algorithm to solve (\ref{NN_training_stochastic}), using   
\begin{align}\label{NN_training_surrogate}
\widetilde{f}_{i}\left(\mathbf{w}_{i};\mathbf{w}_{i}^t,\boldsymbol{\xi}^t\right)\,=\, \frac{1}{|{\cal B}_i^t|}\sum_{m\in {\cal B}_i^t} l\left(y_{i,m},\widetilde{g}(\mathbf{w}_i;\mathbf{w}_i^t,\mathbf{x}_{i,m})\right) +\frac{\tau}{2}\lVert \mathbf{w} - \mathbf{w}_i^t \rVert^2
\end{align}
as a local strongly convex surrogate function for $f_i(\mathbf{w},\boldsymbol{\xi})$, where $\tau>0$, and 
$$\widetilde{g}(\mathbf{w};\mathbf{w}_i^t,\mathbf{x}_{i,m})=g(\mathbf{w}_i^t,\mathbf{x}_{i,m})+\nabla g(\mathbf{w}_i^t,\mathbf{x}_{i,m})^T(\mathbf{w}_i-\mathbf{w}_i^t)$$
represents the linearization of the NN function $g$ around $\mathbf{w}_i^t$, for a given input $\mathbf{x}_{i,m}$. The terms $\nabla g(\mathbf{w}_i^t,\mathbf{x}_{i,m})$, for all $i$ and $m$, can be computed by standard back-propagation \cite{wilamowski2008computing}.

\noindent \textbf{A practical example.} Consider $l(a,b)=(a-b)^2$ and $G(\mathbf{w})=\lambda\|\mathbf{w}\|^2$ in (\ref{NN_training_stochastic}). This setting is largely used in regression type problems. Now, letting $\mathbf{J}_{i,m}^t=\nabla g(\mathbf{w}_i^t,\mathbf{x}_{i,m})$ and $r_{i,m}^t=y_{i,m}-g(\mathbf{w}_i^t;\mathbf{x}_{i,m})+{\mathbf{J}_{i,m}^t}^T \mathbf{w}_i^t$, we have the closed-form solution for the SCA optimization in Algorithm 1:
\begin{equation}\label{SCA_closed_form}
   \widehat{\mathbf{w}}_i^t= ({\mathbf{A}_i^t})^{-1} \mathbf{b}_i^t, 
\end{equation}
where $\mathbf{A}_i^t$ and $\mathbf{b}_i^t$ write as:
\begin{align}
  & \mathbf{A}_i^t= \frac{\rho^t}{|{\cal B}_i^t|} \sum_{m\in {\cal B}_i^t} {\mathbf{J}_{i,m}^t} {\mathbf{J}_{i,m}^t}^T+\lambda \mathbf{I},\nonumber\\
  & \mathbf{b}_i^t= \frac{\rho^t}{|{\cal B}_i^t|} \sum_{m\in {\cal B}_i^t} {\mathbf{J}_{i,m}^t}r_{i,m}^t - \frac{\rho^t}{2} \widetilde{\boldsymbol{\pi}}_i^t - \frac{(1-\rho^t)}{2} \mathbf{d}_i(t).  \nonumber
\end{align}
The step in (\ref{SCA_closed_form}) is in closed-form, but requires the inversion of a matrix at each iteration. The complexity is of the order of $O(p^3$), where $p$ is the size of the vector $\mathbf{w}$. As shown in \cite{di2016next,scardapane2017framework}, this complexity can be largely reduced exploiting either inexact updates or parallel computation using multiple cores at each network agent. 

%  Each agent is interested in minimizing the standard expected risk $f_i(\mathbf{x}) = \mathbb{E}_{(\mathbf{t}, y) \sim p(t, y)} L(g(\mathbf{x}; \mathbf{t}), y)$, where $L$ is a proper loss function, and the expectation is computed w.r.t. a global probability distribution of the data. In practice, each agent is endowed with a local dataset of examples. At every iteration of optimization, each agent draws a mini-batch of $B_i$ elements $\mathcal{B}_i^t = (\mathbf{t}_{i,m}, y_{i,m})_{m=1}^{B_i}$ for approximating the global expected risk.

% To build the surrogate function, one could simply linearize the local cost functions, a strategy that we call \textit{full linearization} (FL). However, similarly to \cite{scardapane2017framework}, we can exploit the knowledge of the optimization problem to build a more sophisticated surrogate function, by only linearizing the network model $g$, leading to a strategy we call \textit{partial linearization} (PL):
% %
% \begin{equation}
%     \widetilde{g}(\mathbf{x}; \mathbf{x}_i^t, \mathbf{t}_{i,m}) = g(\mathbf{x}_i^t; \mathbf{t}_{i,m}) + \left(\mathbf{J}_{i,m}^t\right)^T\left(\mathbf{x} - \mathbf{x}_i^t\right) \,,
%     \label{eq:linearized_neural_network}
% \end{equation}
% %
% where $\mathbf{J}_{i,m}^t = \nabla g(\mathbf{x}_i^t; \mathbf{t}_{i,m})$ can be computed by standard back-propagation \cite{wilamowski2008computing}.

\section{Numerical Results}

In this section, we evaluate numerically the performance of the S-NEXT algorithm, considering the practical case elaborated in the previous Section (i.e., the update in (\ref{SCA_closed_form})), and applying it to two regression problems: the Boston dataset\footnote{\url{https://www.cs.toronto.edu/~delve/data/boston/bostonDetail.html}}, and the SML2010 dataset\footnote{\url{https://archive.ics.uci.edu/ml/datasets/SML2010}}. In both cases, we randomly distribute the training examples on a connected network composed of $6$ agents and having random topology. Regarding the neural network architecture, we consider models with two hidden layers having $30$ units each, and $\tanh$ non-linearities. In Fig. 1, we compare the learning rate of S-NEXT with a distributed stochastic gradient descent (SGD) procedure \cite{vlaski2019distributed}. We also consider three additional strong baselines, which are centralized implementations of SGD, SCA \cite{yang2016parallel}, and Adam \cite{kingma2014adam}. In these centralized cases, we assume all training data available at a single processing unit that performs standard optimization with these baselines. The regularization parameter is set to $\lambda = 10^{-2}$. The parameter setting for S-NEXT considers a time varying step-size rule given by $\alpha^t=\alpha^{t-1}(1-\varepsilon \alpha^{t-1})$, having initial learning rate of $\alpha^0 = 0.01$, and decaying factor $\varepsilon=10^{-3}$; similarly, $\rho^t$ follows the same decaying rule with $\rho^0 = 0.9$ and $\varepsilon=5\times10^{-4}$. The learning rates and the hyper-parameters of all other methods are fine-tuned to provide the fastest convergence behavior. We implement the simulation in the JAX framework \cite{jax2018github}.\footnote{\url{https://jax.readthedocs.io/}} As we can notice from Fig. 1, the proposed method converges to the solution of its centralized counterpart, i.e., the stochastic SCA method \cite{yang2016parallel}, which is provably convergent to stationary solutions of problem (\ref{NN_training_stochastic}). This confirms the theoretical results of Theorem 2. Furthermore, when compared to distributed SGD, our method illustrates a much faster convergence behavior to generally better locally optimal solutions of (\ref{NN_training_stochastic}). Interestingly, S-NEXT outperforms also the centralized Adam algorithm from \cite{kingma2014adam} in terms of learning rate. These results illustrate the very good performance of the proposed S-NEXT algorithm, when applied to the distributed stochastic training of NN models. 

\begin{figure*}[t]
\subfloat[Boston]{
\includegraphics[width=8cm,keepaspectratio]{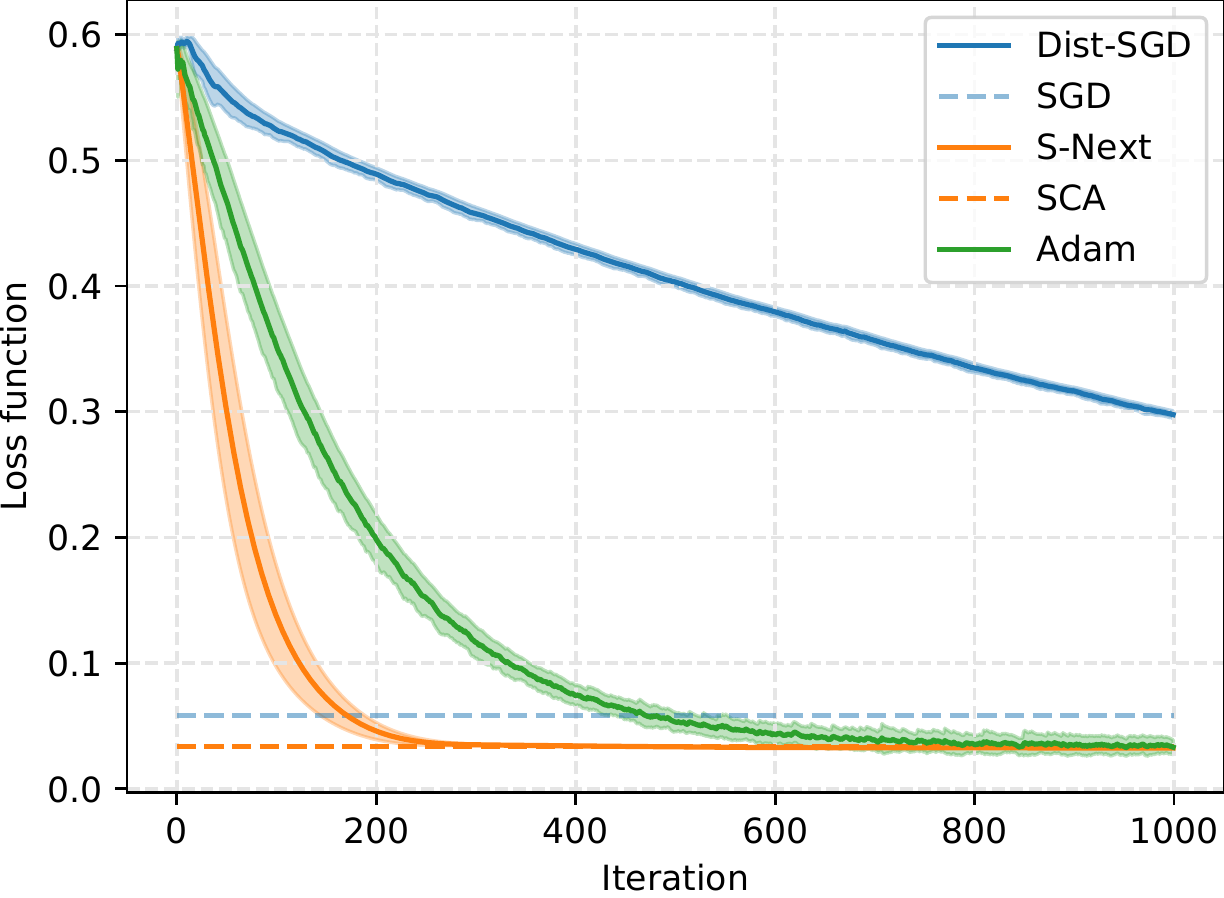}
}
\subfloat[SML2010]{
\includegraphics[width=8cm,keepaspectratio]{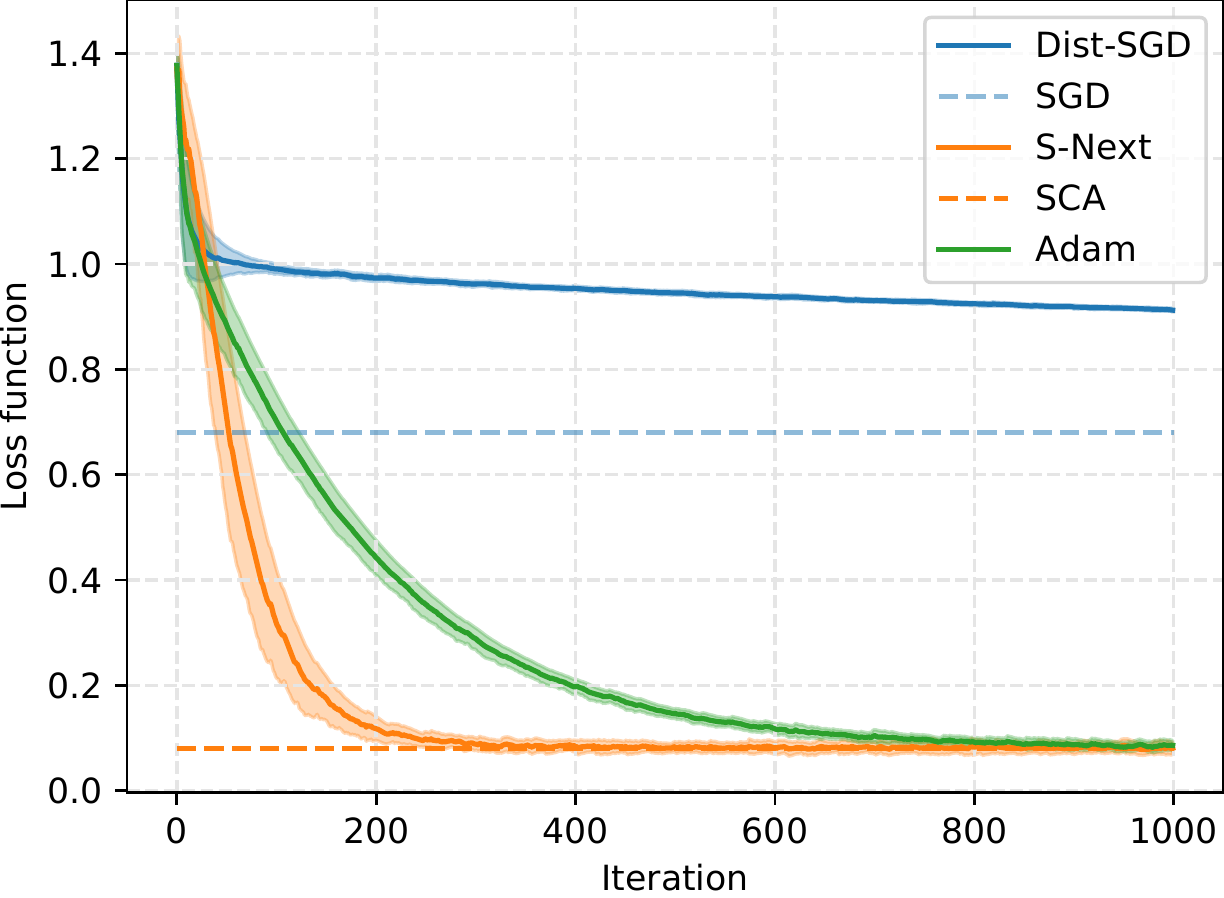}
}
\label{fig:losses_convergence}
\caption{Learning curve of different algorithms applied to prediction tasks}
\end{figure*}

\section{Conclusions}

In this paper we have introduced S-NEXT, a novel algorithmic
framework for stochastic nonconvex distributed optimization in multi-agent networks.  S-NEXT exploits successive convex approximation techniques while leveraging \textit{dynamic} consensus as a gradient tracking mechanism, and recursive average to asymptotically evaluate the expected gradient of the network loss function. Almost sure convergence to (stationary) solutions of the nonconvex problem is established under mild conditions. The proposed method is then customized to the stochastic training of NN models over a multi-agent network. Numerical results confirm the theoretical findings, and show that S-NEXT compares favorably to existing  algorithms for distributed nonconvex stochastic optimization and learning.

\bibliographystyle{MyIEEE}
\bibliography{refs_arxiv}

% Generated by IEEEtran.bst, version: 1.13 (2008/09/30)
\begin{thebibliography}{10}
\providecommand{\url}[1]{#1}
\csname url@samestyle\endcsname
\providecommand{\newblock}{\relax}
\providecommand{\bibinfo}[2]{#2}
\providecommand{\BIBentrySTDinterwordspacing}{\spaceskip=0pt\relax}
\providecommand{\BIBentryALTinterwordstretchfactor}{4}
\providecommand{\BIBentryALTinterwordspacing}{\spaceskip=\fontdimen2\font plus
\BIBentryALTinterwordstretchfactor\fontdimen3\font minus
  \fontdimen4\font\relax}
\providecommand{\BIBforeignlanguage}[2]{{%
\expandafter\ifx\csname l@#1\endcsname\relax
\typeout{** WARNING: IEEEtran.bst: No hyphenation pattern has been}%
\typeout{** loaded for the language `#1'. Using the pattern for}%
\typeout{** the default language instead.}%
\else
\language=\csname l@#1\endcsname
\fi
#2}}
\providecommand{\BIBdecl}{\relax}
\BIBdecl

\bibitem{Nedic-Ozdaglar}
A.~Nedi\'{c} and A.~Ozdaglar, ``Distributed subgradient methods for multiagent
  optimization,'' \emph{IEEE Trans. on Automatic Control}, vol.~54, no.~1, pp.
  48--61, Jan. 2009.

\bibitem{Nedic-Ozdaglar-Parillo}
A.~Nedi\'{c}, A.~Ozdaglar, and P.~Parillo, ``Constrained consensus and
  optimization in multi-agent networks,'' \emph{{IEEE} Trans. on Automatic
  Control}, vol.~55, no.~4, pp. 922--938, 2010.

\bibitem{jakovetic2014fast}
D.~Jakoveti{\'c}, J.~Xavier, and J.~M. Moura, ``Fast distributed gradient
  methods,'' \emph{IEEE Transactions on Automatic Control}, vol.~59, no.~5, pp.
  1131--1146, 2014.

\bibitem{Boyd-Parikh-Chu-Peleato-Eckstein}
S.~Boyd, N.~Parikh, E.~Chu, B.~Peleato, and J.~Eckstein, \emph{Distributed
  optimization and statistical learning via the alternating direction method of
  multipliers}, ser. Foundations and Trends in Machine Learning.\hskip 1em plus
  0.5em minus 0.4em\relax Boston-Delft: NOW Publishers, 2011, vol.~3, no.~1.

\bibitem{Cattivelli-Sayed}
F.~S. Cattivelli and A.~H. Sayed, ``Diffusion {LMS} strategies for distributed
  estimation,'' \emph{IEEE Trans. on Signal Processing}, vol.~58, pp.
  1035--1048, March 2010.

\bibitem{Chen-Sayed}
J.~Chen and A.~H. Sayed, ``Diffusion adaptation strategies for distributed
  optimization and learning over networks,'' \emph{IEEE Trans. on Signal
  Processing}, vol.~60, no.~8, pp. 4289--4305, August 2012.

\bibitem{DiLo-Sayed}
P.~{Di Lorenzo} and A.~H. Sayed, ``Sparse distributed learning based on
  diffusion adaptation,'' \emph{IEEE Trans. on Signal Processing}, vol.~61,
  no.~6, pp. 1419--1433, March 2013.

\bibitem{vlaski2016diffusion}
S.~Vlaski, L.~Vandenberghe, and A.~H. Sayed, ``Diffusion stochastic
  optimization with non-smooth regularizers,'' in \emph{2016 IEEE International
  Conference on Acoustics, Speech and Signal Processing (ICASSP)}.\hskip 1em
  plus 0.5em minus 0.4em\relax IEEE, 2016, pp. 4149--4153.

\bibitem{ouyang2013stochastic}
H.~Ouyang, N.~He, L.~Tran, and A.~Gray, ``Stochastic alternating direction
  method of multipliers,'' in \emph{International Conference on Machine
  Learning}, 2013, pp. 80--88.

\bibitem{zhong2014fast}
W.~Zhong and J.~Kwok, ``Fast stochastic alternating direction method of
  multipliers,'' in \emph{International Conference on Machine Learning}, 2014,
  pp. 46--54.

\bibitem{zeng2018nonconvex}
J.~Zeng and W.~Yin, ``On nonconvex decentralized gradient descent,'' \emph{IEEE
  Trans. on Signal Processing}, vol.~66, no.~11, pp. 2834--2848, 2018.

\bibitem{wai2017decentralized}
H.-T. Wai, J.~Lafond, A.~Scaglione, and E.~Moulines, ``Decentralized
  {Frank--Wolfe} algorithm for convex and nonconvex problems,'' \emph{IEEE
  Trans. on Automatic Control}, vol.~62, no.~11, pp. 5522--5537, 2017.

\bibitem{di2016next}
P.~Di~Lorenzo and G.~Scutari, ``Next: In-network nonconvex optimization,''
  \emph{IEEE Transactions on Signal and Information Processing over Networks},
  vol.~2, no.~2, pp. 120--136, 2016.

\bibitem{hong2017prox}
M.~Hong, D.~Hajinezhad, and M.-M. Zhao, ``Prox-{PDA}: The proximal primal-dual
  algorithm for fast distributed nonconvex optimization and learning over
  networks,'' in \emph{Proc. of ICML}, 2017, pp. 1529--1538.

\bibitem{swenson2019annealing}
B.~Swenson, S.~Kar, H.~V. Poor, and J.~Moura, ``Annealing for distributed
  global optimization,'' \emph{arXiv preprint arXiv:1903.07258}, 2019.

\bibitem{vlaski2019distributed}
S.~Vlaski and A.~H. Sayed, ``Distributed learning in non-convex
  environments--part i: Agreement at a linear rate,'' \emph{arXiv preprint
  arXiv:1907.01848}, 2019.

\bibitem{vlaski2019distributed2}
------, ``Distributed learning in non-convex environments--part ii: Polynomial
  escape from saddle-points,'' \emph{arXiv preprint arXiv:1907.01849}, 2019.

\bibitem{vlaski2019second}
------, ``Second-order guarantees of stochastic gradient descent in non-convex
  optimization,'' \emph{arXiv preprint arXiv:1908.07023}, 2019.

\bibitem{george2019distributed}
J.~George, T.~Yang, H.~Bai, and P.~Gurram, ``Distributed stochastic gradient
  method for non-convex problems with applications in supervised learning,''
  \emph{arXiv preprint arXiv:1908.06693}, 2019.

\bibitem{lu2019gnsd}
S.~Lu, X.~Zhang, H.~Sun, and M.~Hong, ``Gnsd: a gradient-tracking based
  nonconvex stochastic algorithm for decentralized optimization,'' in
  \emph{Proceedings of IEEE Data Science Workshop (DSW)}, 2019, pp. 315--321.

\bibitem{Scutari-Facchinei-Song-Palomar-Pang}
G.~Scutari, F.~Facchinei, P.~Song, D.~P. Palomar, and J.-S. Pang,
  ``Decomposition by partial linearization: Parallel optimization of multiuser
  systems,'' \emph{IEEE Trans. on Signal Processing}, vol.~63, no.~3, pp.
  641--656, Feb. 2014.

\bibitem{Scutari-Facchinei-Sagratella}
F.~Facchinei, G.~Scutari, and S.~Sagratella, ``Parallel selective algorithms
  for nonconvex big data optimization,'' \emph{IEEE Trans. on Signal
  Processing}, vol.~63, no.~7, pp. 1874--1889, April 2015.

\bibitem{yang2016parallel}
Y.~Yang, G.~Scutari, D.~P. Palomar, and M.~Pesavento, ``A parallel
  decomposition method for nonconvex stochastic multi-agent optimization
  problems,'' \emph{IEEE Transactions on Signal Processing}, vol.~64, no.~11,
  pp. 2949--2964, 2016.

\bibitem{Zhu-Martinez}
M.~Zhu and S.~Mart\'{\i}nez, ``Discrete-time dynamic average consensus,''
  \emph{Automatica}, vol.~46, no.~2, pp. 322--329, Feb. 2010.

\bibitem{scardapane2017framework}
S.~Scardapane and P.~Di~Lorenzo, ``A framework for parallel and distributed
  training of neural networks,'' \emph{Neural Networks}, vol.~91, pp. 42--54,
  2017.

\bibitem{wen2017terngrad}
W.~Wen, C.~Xu, F.~Yan, C.~Wu, Y.~Wang, Y.~Chen, and H.~Li, ``Terngrad: Ternary
  gradients to reduce communication in distributed deep learning,'' in
  \emph{Advances in neural information processing systems}, 2017, pp.
  1509--1519.

\bibitem{wilamowski2008computing}
B.~M. Wilamowski, N.~J. Cotton, O.~Kaynak, and G.~Dundar, ``Computing gradient
  vector and jacobian matrix in arbitrarily connected neural networks,''
  \emph{IEEE Transactions on Industrial Electronics}, vol.~55, no.~10, pp.
  3784--3790, 2008.

\bibitem{kingma2014adam}
D.~P. Kingma and J.~Ba, ``Adam: A method for stochastic optimization,''
  \emph{arXiv preprint arXiv:1412.6980}, 2014.

\bibitem{jax2018github}
\BIBentryALTinterwordspacing
J.~Bradbury, R.~Frostig, P.~Hawkins, M.~J. Johnson, C.~Leary, D.~Maclaurin, and
  S.~Wanderman-Milne, ``{JAX}: composable transformations of {P}ython+{N}um{P}y
  programs,'' 2018.
\BIBentrySTDinterwordspacing

\end{thebibliography}

\end{document}